\documentclass[conference]{IEEEtran}
\IEEEoverridecommandlockouts

\usepackage{cite}

\usepackage{amsmath,amssymb,amsfonts}
\usepackage{amsthm}
\usepackage[linesnumbered,ruled,vlined]{algorithm2e} 

\usepackage{graphicx}
\usepackage{textcomp}
\usepackage{xcolor}
\usepackage{subfigure}
\usepackage{multirow}
\usepackage{mathtools}
\usepackage{hyperref}
\usepackage{booktabs}

\allowdisplaybreaks

\def\BibTeX{{\rm B\kern-.05em{\sc i\kern-.025em b}\kern-.08em
    T\kern-.1667em\lower.7ex\hbox{E}\kern-.125emX}}
\begin{document}

\makeatletter
\newcommand{\linebreakand}{%
  \end{@IEEEauthorhalign}
  \hfill\mbox{}\par
  \mbox{}\hfill\begin{@IEEEauthorhalign}
}
\makeatother

\title{A Feasibility-Preserved Quantum Approximate Solver for the Capacitated Vehicle Routing Problem}

\author{\IEEEauthorblockN{1\textsuperscript{st} Ningyi Xie}
    \IEEEauthorblockA{\textit{Graduate School of Science and Technology} \\
    \textit{University of Tsukuba}\\
    Ibaraki Prefecture, Japan \\
    nyxie@cavelab.cs.tsukuba.ac.jp}
    \and
    \IEEEauthorblockN{2\textsuperscript{nd} Xinwei Lee}
    \IEEEauthorblockA{\textit{School of Computing and Information Systems} \\
    \textit{Singapore Management University}\\
    Ibaraki Prefecture, Japan \\
    xwlee@smu.edu.sg}
    \linebreakand 
    \IEEEauthorblockN{3\textsuperscript{rd} Dongsheng Cai}
    \IEEEauthorblockA{\textit{Faculty of Engineering, Information and Systems} \\
    \textit{University of Tsukuba}\\
    Ibaraki Prefecture, Japan \\
    cai@cs.tsukuba.ac.jp}
    \and
    \IEEEauthorblockN{4\textsuperscript{th} Yoshiyuki Saito}
    \IEEEauthorblockA{\textit{Graduate School of Computer Science and Engineering} \\
    \textit{University of Aizu}\\
    Fukushima Prefecture, Japan \\
    d8241104@u-aizu.ac.jp}
    \linebreakand 
    \IEEEauthorblockN{5\textsuperscript{th} Nobuyoshi Asai}
    \IEEEauthorblockA{\textit{School of Computer Science and Engineering} \\
    \textit{University of Aizu}\\
    Fukushima Prefecture, Japan \\
    nobuyoshi.asai@gmail.com}
    \and
    \IEEEauthorblockN{6\textsuperscript{th} Hoong Chuin Lau}
    \IEEEauthorblockA{\textit{School of Computing and Information Systems} \\
    \textit{Singapore Management University}\\
    80 Stamford Road, Singapore \\
    hclau@smu.edu.sg}
    }

\maketitle

\begin{abstract}
    The Capacitated Vehicle Routing Problem (CVRP) is an NP-optimization problem (NPO) that arises in various fields including transportation and logistics.
    The CVRP extends from the Vehicle Routing Problem (VRP),
    aiming to determine the most efficient plan for a fleet of vehicles to deliver goods to a set of customers,
    subject to the limited carrying capacity of each vehicle.
    As the number of possible solutions skyrockets when the number of customers increases,
    finding the optimal solution remains a significant challenge.
    Recently, the Quantum Approximate Optimization Algorithm (QAOA), 
    a quantum-classical hybrid algorithm, 
    has exhibited enhanced performance in certain combinatorial optimization problems compared to classical heuristics. 
    However, its ability diminishes notably in solving constrained optimization problems including the CVRP. 
    This limitation primarily arises from the typical approach of encoding the given problems as penalty-inclusive binary optimization problems. 
    In this case, the QAOA faces challenges in sampling solutions satisfying all constraints. 
    Addressing this, our work presents a new binary encoding for the CVRP, 
    with an alternative objective function of minimizing the shortest path that bypasses the vehicle capacity constraint of the CVRP. 
    The search space is further restricted by the constraint-preserving mixing operation. 
    We examine and discuss the effectiveness of the proposed encoding under the framework of the variant of the QAOA, 
    Quantum Alternating Operator Ansatz (AOA), through its application to several illustrative examples.
    Compared to the typical QAOA approach, 
    the proposed method not only preserves the feasibility but also achieves a significant enhancement in the probability of measuring optimal solutions.
\end{abstract}

\section{Introduction}

\begin{figure}[t]
    \centering
    \includegraphics[width=60mm]{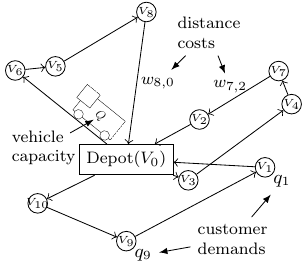}
    \caption{Visualization of the CVRP.}\label{fig:cvrp}
\end{figure}

The Capacitated Vehicle Routing Problem (CVRP) is an NP-hard combinatorial optimization problem attracted significant attention
from both industry and science due to its broad applicability and inherent complexity, since its proposal in \cite{dantzig1959truck}.
The objective of the CVRP is to find an optimal set of routes for a fleet of vehicles to serve a set of customers,
seeking to minimize the time or total distance, etc.
The problem can be defined on a graph $G = (V, E)$, as shown in Figure 1. 
For an $N$-customer CVRP, 
we define $V \!= \!\{V_0,V_1,\dots,V_{N}\}$ to be the set of vertices, 
with the vertex $V_0$ as the depot,
and vertex $V_i$, $i\!\in\!\{1,2,\dots,N\}$, signifies the location of the customer $i$ with a certain demand $q_i$.
Here, $E$ is defined to be a set of directed edges.
Each edge $(i,j) \!\in \!E$ has a weight $w_{ij}$, indicating the distance for traveling from customer $i$ to customer $j$.
With vehicles having a capacity $Q$ $(Q\!\geqslant \!q_i)$, the solution comprises a set of routes that meet the following constraints:
\begin{itemize}
    \item Depot Constraint: All routes must start and end at the depot;
    \item Customer Visit Constraint: Each customer must be visited exactly once by one of the vehicles;
    \item Vehicle Capacity Constraint: The total demand of customers along a route must not exceed the capacity of the vehicle.
\end{itemize}
This paper presents a quantum heuristic method focuses on achieving the solution with the shortest total distance.

Recently, there has been significant progress in quantum computing.
A plethora of quantum algorithms has been proposed \cite{grover1996fast,shor1999polynomial,montanaro2016quantum},
aiming to achieve computational advantages over their classical counterparts.
One of them in solving combinatorial optimization problems is the Quantum Approximate Optimization Algorithm (QAOA) \cite{farhi2014quantum}
or its variant the Quantum Alternating Operator Ansatz (AOA) \cite{hadfield2019quantum}.

Given a binary assignment combinatorial problem $(F,C)$,
the objective is to find the optimal solution $x^{\ast}\! \in F\! \subseteq \{0,1\}^n $, corresponding to the minimum of $C(x)$,
where $n$ denotes the number of required decision bits.
The QAOA encodes the solution,
using an $n$-qubit parameterized state that evolves from the ground state of the initial Hamiltonian ($H_M$) through a trotterized adiabatic time evolution in $p$ steps.
The parameters are optimized to force the prepared state to approach the ground state of the cost-based Hamiltonian ($H_C$),
which encodes the optimal solution to the given problem. Thus, the QAOA gives the approximate solution for the combinatorial optimization problems.

The advantage of the QAOA to solve unconstrained combinatorial problems, such as Max-Cut, is discussed in \cite{crooks2018performance,moussa2020quantum}.
However, the QAOA solutions suffer from a low feasible solution ratio for constrained problems whose $F\subsetneqq  \{0,1\}^n$,
with several experimental studies illustrating this limitation even in toy examples \cite{azad2022solving,glos2022space,awasthi2023quantum,palackal2023quantum}.
Since the penalty terms are incorporated for infeasible solutions, 
as shown in Figure~\ref{fig:fig1a}, 
the energy landscape produced by QAOA has many local minima and barren plateaus in the typical parameter range.

Instead of penalizing the infeasible solutions, 
the variant AOA aims to preserve the feasibility by specifying a mixing operation according to the constraints. 
Additionally, recent studies \cite{wang2020x,cook2020quantum} suggest that initializing the state as an equal superposition of all feasible states could be beneficial. 
Figure~\ref{fig:fig1b} shows that the energy landscape becomes smoother and more conducive to optimization when the ansatz is initialized 
from an equal superposition and further employs the Grover Mixer \cite{bartschi2020grover} to preserve the feasibility. 
However, either finding constraint-preserving mixers or identifying operations for preparing such initial states present challenges, 
which limits the range of applicability for the AOA. This limitation extends to the currently binary-formulated CVRPs \cite{palackal2023quantum,feld2019hybrid}.

\begin{figure}[th]
    \centering     
    \subfigure[QAOA]{\label{fig:fig1a}\includegraphics[width=40mm]{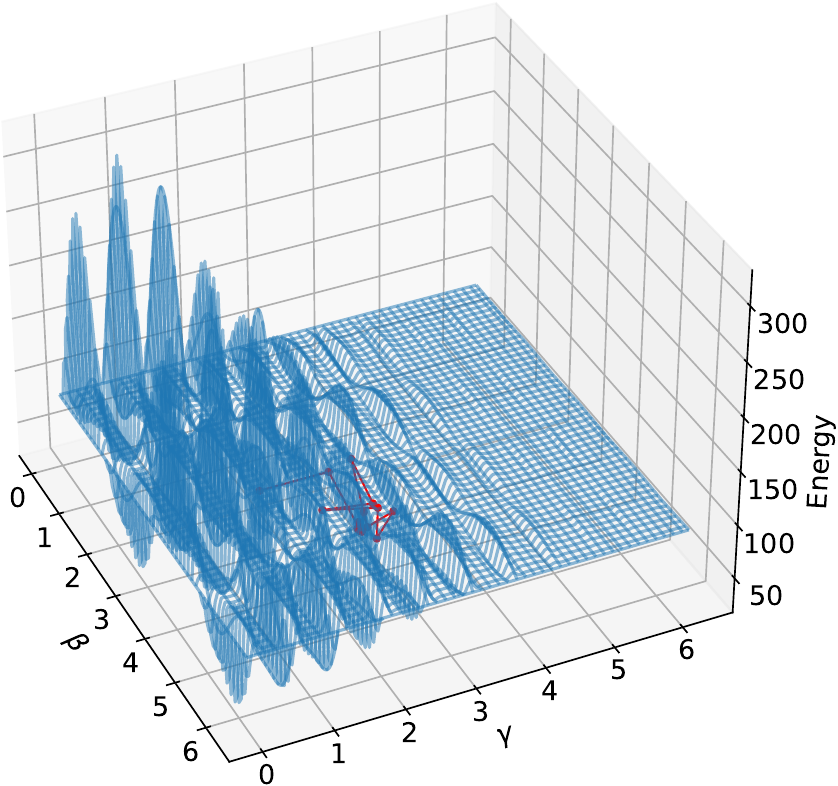}}
    \subfigure[GM-QAOA]{\label{fig:fig1b}\includegraphics[width=40mm]{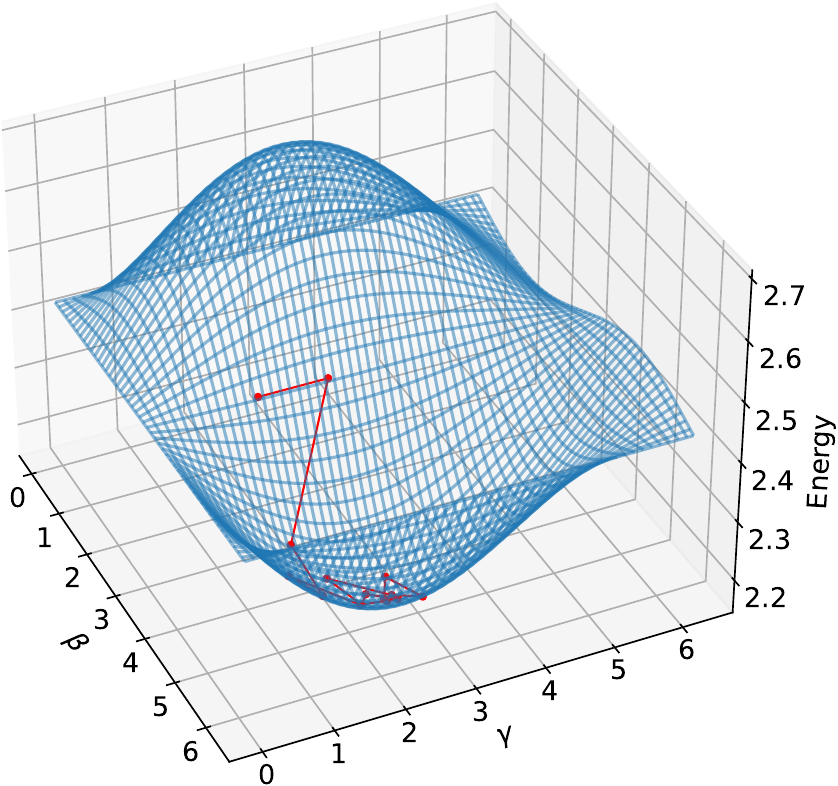}}
    \caption{The energy landscapes of depth-$1$ \subref{fig:fig1a} QAOA and \subref{fig:fig1b} Grover-Mixer Quantum Alternating Operator Ansatz (GM-QAOA) for solving the Traveling Salesman Problem (TSP).
        The red lines track the optimization process.
        The optimizer finds the position of the lowest energy state in the GM-QAOA,
        while it becomes trapped in a local minimum within the QAOA.}\label{fig:landscape}
\end{figure}

In this work, we explore the application of the AOA for solving the CVRP\footnote{The implementation is available at
\href{https://github.com/NingyiXie/GM\_QAOA\_for\_CVRP}{https://github.com/NingyiXie/GM\_QA OA\_for\_CVRP}}.
We propose an $\mathcal{O}(N^2)$ binary variable encoding of the solution. 
This problem encoding enables the existence of constraint-preserving mixers that restrict the customer visit constraint, 
as well as the operation to prepare a uniform superposition for initializing. 
Concurrently, the solution decoding process is conditionally executed to bypass the vehicle capacity constraint. 
Considering the cost Hamiltonian also becomes condition-dependent in this framework, 
we propose an auxiliary circuit for condition encoding. Incorporating this with other operations, 
the quantum circuit comprises $\mathcal{O}(pN^2(N+(\log_2Q)^2))$ gates. 
Besides, the encoding inherently satisfies the depot constraint, 
thus, the AOA solver we developed effectively preserves the feasibility. 
Additionally, the AOA states result in a higher probability of measuring optimal states, 
as evidenced by our experimental results.

The paper is organized as follows. 
Section \ref{sec:aoa} gives a background of the AOA. 
Section \ref{sec:related_work} describes previous research on quantum solutions for the CVRP. 
The proposed method is described in Section \ref{sec:the_method} and numerically analyzed in Section \ref{sec:experiments}. 
Finally, we conclude this paper in Section \ref{sec:conclusion}.

\section{The Quantum Alternating Operator Ansatz}
\label{sec:aoa}
In the QAOA and its variants \cite{farhi2014quantum,hadfield2019quantum,wang2020x,bartschi2020grover},
the initial step typically involves the encoding of the given problem into a binary assignment combinatorial problem, 
denoted as $(F, C)$. When $F\subsetneqq  \{0,1\}^n$, the problem is constrained. 
For the encoded combinatorial problem with $a$ equality constraints and $b$ inequality constraints, 
\begin{equation}
    \begin{aligned}
    & \min_{x\in\{0,1\}^n} 
    & & C(x) \\
    & \text{subject to}
    & & h_j(x) = 0, \; i = 1, \ldots, a, \\
    &&& g_i(x) \leq 0, \; j = 1, \ldots, b,
    \end{aligned}
\end{equation}
the QAOA typically converts the problem into a quadratic unconstrained binary optimization (QUBO) problem \cite{boros2007local}
or a higher-order binary optimization (HOBO) problem \cite{mandal2020compressed}, by combining the cost function and penalty terms in Lagrange form, as follows, 
\begin{equation}
    \begin{aligned}
        & \quad \quad \min_{x,y\in\{0,1\}^{n+m}} \mathcal{C}(x,y),\\
        & \mathcal{C}(x,y) = C(x)+\sum_{i=1}^{a}\lambda_i h_i(x)^2+\sum_{i=1}^{b}\mu_i(g_i(x)+\sum_{j=0}2^j y_{ij})^2
    \end{aligned}
    \label{eqn:penalized_obj}
\end{equation}
where multipliers $\lambda$ and $\mu$ are predefined. The variable $y$ contains $m$ bits, serving as slack variables.

In contrast, the search space in the AOA is ideally restricted to the feasible set $F$, thereby negating the need for further transformation as,
\begin{equation}
    \min_{x\in F} C(x).
\end{equation}

Inspired by adiabatic quantum computing \cite{farhi2000quantum},
the QAOAs prepare a parameterized state denoted by $ \vert \psi_p(\boldsymbol{\gamma},\boldsymbol{\beta}) \rangle $.
This state evolves from an initial state $\vert \psi_0 \rangle$ through a series of $p$ repetitions,
wherein two distinctive types of operations are employed, the phase separation operation $U_{P}$ and the mixing operation $U_{M}$,
as the following formulation,
\begin{align}
    \vert \psi_p(\boldsymbol{\gamma},\boldsymbol{\beta}) \rangle \coloneqq \prod_{j=1}^{p} U_{M}(\beta_{j})U_{P}(\gamma_{j})  \vert \psi_0 \rangle.
\end{align}
\begin{itemize}
    \item The phase separation, $U_P$, encodes the given problem, functioning such that $U_P(\gamma)\vert x \rangle \!=\!e^{-i\gamma C(x)}\vert x \rangle$. 
    Generally, it is defined as $U_P(\gamma)\! \coloneqq \! e^{-i\gamma H_C}$, where cost Hamiltonian $H_C$ satisfies $\langle x\vert H_{C} \vert x\rangle\!=\!C(x)$. 
    $H_C$ is constructed by replacing binary variables with $(I\!-\!Z_i)/2$, where $Z_i$ denotes the Pauli-$Z$ operation acting on the $i^{th}$ qubit.
    \item $ U_M(\beta)\! \coloneqq \!e^{-i\beta H_M} $ facilitates transitions between different quantum states in search space. 
    As the AOA necessitates that $U_M$ preserves the feasible subspace, 
    the $H_M$ is tailored according to the feasible set $F$, ensuring that 
    \begin{equation}
        \langle x'\vert H_M \vert x \rangle=0,\; x \notin F \oplus x' \notin F.
    \end{equation}
    The mixing operation for the proposed CVRP encoding is discussed in Section \ref{sec:mixer}.
    \item The initial state $\vert \psi_0 \rangle$ can be any state that encodes feasible solutions. 
    However, experimental results in \cite{wang2020x,cook2020quantum} suggest that an equal amplitude superposition of all feasible states as, 
    \begin{equation}
        \vert \psi_0 \rangle = \vert F \rangle \coloneqq U_S \vert 0 \rangle ^{\otimes n} = \frac{1}{\sqrt{\vert F \vert }}\sum_{x \in F} \vert x\rangle,
    \end{equation}
    may be more advantageous compared to a random selection, where $U_S$ is the preparation operation. 
    At present, $U_S$ has been polynomially formulated for the unconstrained problems,
    the permutation-based problems (i.e., problems involving the ordering of elements),
    and the problems whose feasible set $F$ exactly contains solutions with the same Hamming weight \cite{bartschi2020grover}.
\end{itemize}
The depth $p$ is pre-defined.
$\boldsymbol{\gamma}\! =\! [\gamma_1,\gamma_2,\dots,\gamma_p]$ and $\boldsymbol{\beta} \!= \![\beta_1,\beta_2,\dots,\beta_p]$ are trainable parameters.
They are tuned to find the ground state of $H_{C}$ by optimizing the expected value $E_p(\boldsymbol{\gamma},\boldsymbol{\beta})$:
\begin{align}
    \label{eqn:expect}
    E_p(\boldsymbol{\gamma},\boldsymbol{\beta}) \coloneqq \langle\psi_p(\boldsymbol{\gamma},\boldsymbol{\beta}) \vert H_C \vert \psi_p(\boldsymbol{\gamma},\boldsymbol{\beta}) \rangle.
\end{align}

\section{Related Work}
\label{sec:related_work}
Recently, several QUBO formulations have been proposed for the CVRP and its variants. 
The QUBO formulations for the CVRP are derived in \cite{feld2019hybrid,palackal2023quantum}.
The solutions are encoded using $T\times N\times K$ decision bits, 
where $T$, $N$, and $K$, respectively, represent the number of time steps, customers, and vehicles. 
Both introduce additional qubits as slack variables, 
while \cite{palackal2023quantum} uses a logarithmic encoding as same to Equation (\ref{eqn:penalized_obj}), 
which requires fewer qubits. 
As the vehicle number $K$ is predefined, the search space may not cover the optimal solution in some cases. 
\cite{irie2019quantum} adopts a similar encoding strategy, presenting a QUBO formulation for the CVRP with Time Windows. 
In \cite{harikrishnakumar2020quantum,azad2022solving}, the binary variables are implemented to represent the decisions associated with the solution edges of the VRP and the Multi-depot CVRP.

An alternative approach to the CVRP lies in the 2-Phase-Heuristic \cite{laporte2002classical}.
In the first phase, customers are clustered into several groups such that the total demand of each group does not surpass the vehicle's capacity.
The second phase addresses the Traveling Salesman Problem (TSP) within each group.
\cite{feld2019hybrid} propose a hybrid solution that adopts a classical algorithm \cite{shin2011centroid} for the clustering phase,
while the remaining TSPs are solved using quantum annealing \cite{de2011introduction}.
Both phases can be converted into QUBO forms.
Specifically, the clustering phase is an adaptation of the Multiple Knapsack problem (MKP).
The QUBO forms of the MKP and the TSP are discussed in detail in \cite{lucas2014ising, feld2019hybrid, ruan2020quantum,  glos2022space, palackal2023quantum,awasthi2023quantum}.

Both $1$-Phase and $2$-Phase (i.e., TSP and MKP) QUBO formulations for solving the CVRP include penalty terms. 
The use of the QAOA to solve such QUBOs is anticipated to yield poor-quality results due to the intricate energy landscape. 
Meanwhile, 
as the $1$-Phase CVRP encoding \cite{feld2019hybrid,palackal2023quantum} and MKP encoding \cite{awasthi2023quantum} 
lack constraint-preserving mixers for their feasible sets to the best of our knowledge, the AOA is not applicable.
To mitigate these limitations, this study introduces a new formulation for the CVRP. 
Herein, the solution encoding is divided into two parts: the first represents a permutation of customers while the second remains unconstrained, 
serving to transform the routes into cyclical paths. 
Employing the mixers outlined in \cite{hadfield2019quantum,bartschi2020grover} for the permutation matrix encoding, 
the AOA can be applied to solve the CVRP. 
Notably, our encoding does not need a predetermination of vehicle number, 
which expands the search space but with fewer decision bits.

\begin{algorithm}[h]
    \caption{Solution decoding}
    \label{alg:algorithm}
    \KwIn{Encoding $x,y$; Demands $q$; Capacity $Q$}
    \KwOut{Solution $s$}
    
    $i \gets \arg\max x_1$\;
    $s \gets [(0,i)]$\;
    $d \gets q_{i}$\;
    \For{$t \sim  \left[2,3,\dots,N\right]$}{
        $j \gets \arg\max x_t$\;
        \If{$d+q_{j}\leqslant Q$ and $y_t = 0$}{
            append $(i, j)$ to $s$\;
            $d \gets d+q_{j}$\;
        }
        \Else{
            append $(i, 0)$ to $s$\;
            append $(0, j)$ to $s$\;
            $d \gets q_{j}$\;
        }
        $i \gets j$\;
    }
    append $(i, 0)$ to $s$\;
    \Return $s$\;
\end{algorithm}

\section{AOA for CVRP}
\label{sec:the_method}
In this section, we describe the proposed problem encoding and the structure of the quantum circuit.

\subsection{The Problem Encoding}
\label{sec:the_problem_encoding}
The CVRP can be conceptualized as planning the shortest singular route undertaken by one vehicle,
permitted to refill its capacity at the depot.
While the solution is identified as the collection of sub-routes,
each starting and ending at the depot.
In this scenario,
customers are served sequentially,
whereby the order in which customers are visited corresponds to a permutation.
For an $N$-customer CVRP defined on graph $G\!=\!(V,E)$, 
where $ \vert V \vert\! = \!N\!+\!1 $,
a one-hot vector $x_t\! \in\! \{0,1\}^{N}$, 
subject to $ \sum_{i=1}^{N}x_{t,i}\!=\!1$, 
is employed to represent the customer being served at time step $t \in \{1,2,\dots,N\}$.
$ x_{t,i}=1$ signifies that the customer $i$ is visited at time step $t$.
Furthermore, considering that each customer is visited exactly once,
these one-hot vectors encode different customer numbers.
Thus, the feasible space of $x$ is derived as follows:
\begin{subequations}
    \begin{align}
        F_x \coloneqq \{ & x \vert \, x\in\{0,1\}^{N\times N}; \label{eqn:fxcons1}                     \\
                & \forall\,t\in\{1,2,\dots,N\},\, \sum_{i=1}^{N}x_{t,i}=1; \label{eqn:fxcons2}   \\
                & \forall\,i\in\{1,2,\dots,N\},\, \sum_{t=1}^{N}x_{t,i}=1\}, \label{eqn:fxcons3}
    \end{align}
    \label{eqn:fx}
\end{subequations}
where $x$ is defined as a permutation matrix.

The vehicle is granted the option to return to the depot before reaching the customer at time steps ranging from $2$ to $N$.
This decision is encoded through another binary string $y\!\coloneqq\!y_2 y_3 \dots y_{N}$,
with $y_t\!=\!1$ indicating a depot visit between time steps $t\!-\!1$ and $t$.
In addition, a depot visit also arises when the current capacity fails to satisfy the demand of the next customer.
To address these possibilities, we introduce conditional functions within the process of decoding a solution,
as summarized in Algorithm~\ref{alg:algorithm},
and further elucidated with an example illustrated in Figure~\ref{fig:problem_encoding}.
Thus, $y$ enjoys an unconstrained feasible space, extending over $F_y \coloneqq \{0,1\}^{N-1}$.

Consider the decoding processing (i.e. Algorithm~\ref{alg:algorithm}) as function $D$. 
Given the set of demands q, and vehicle capacity $Q$, 
the cost function of the total distance can be formulated as follows,
\begin{align}
    C(x,y) \coloneqq \sum_{(i,j)\in D(x,y,q,Q)} w_{ij},
\end{align}
where $w_{ij}$ represents the distance traveling from location $V_i$ to location $V_j$.

\begin{figure}[t]
    \centering     
    \subfigure[An example of solution encoding for a $5$-customer CVRP.]{\label{fig:fig2a}\includegraphics[width=42mm]{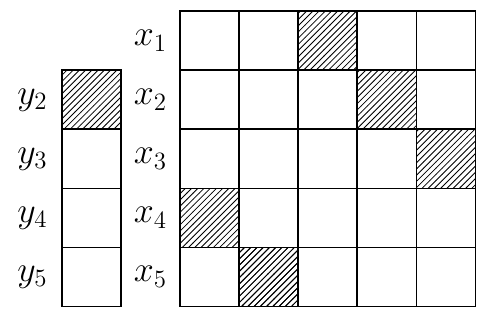}}
    \subfigure[The decoded solution. The node $(V_i, q_i)$ indicates the location of the customer $i$ and the demand $q_i$.]{\label{fig:fig2b}\includegraphics[width=38mm]{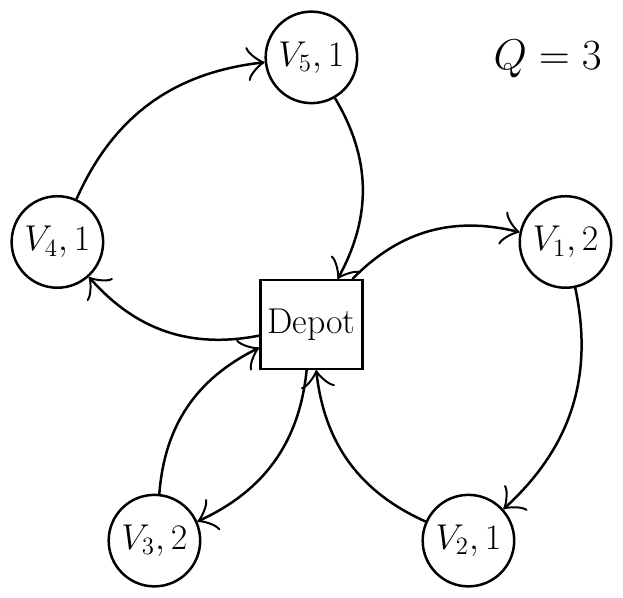}}
    \caption{Visualization of the proposed problem encoding.
        Permutation matrix $x$ exhibit a certain order of customer visits, delineated as $[3,4,5,1,2]$.
        The vehicle returns to the depot before visiting $V_4$ and $V_1$, as decision bit $y_2=1$
        and the insufficiency of the vehicle's capacity at time step $3$ to meet the demand of $V_1$.}\label{fig:problem_encoding}
\end{figure}

\subsection{Preparation Operation}
\label{sec:us}
We employ $N$ numbers of $N$-qubit quantum registers to encode the permutation matrix $x$.
\cite{bartschi2020grover} have introduced a preparation unitary $U_{x}$ for preparing the equal superposition of all possible permutations as:
\begin{align}
    \vert F_{x} \rangle \coloneqq U_{x}\vert 0 \rangle^{\otimes N^2} = \frac{1}{\sqrt{N!}}\sum_{x\in F_x} \vert x \rangle.
\end{align}
This superposition is constructed register by register, involving the following two key steps.
First, the constraint of the register (i.e., (\ref{eqn:fxcons2}) in Equation~(\ref{eqn:fx})) under construction is satisfied applying the $\mathcal{O}(N)$ or $\mathcal{O}(N^2)$ gates controlled by the last register,
while preserving the state of other registers.
Second, the last register is updated to satisfy the column constraint (i.e., (\ref{eqn:fxcons3}) in Equation~(\ref{eqn:fx})).
Hence, the superposition can be prepared by utilizing an $\mathcal{O}(N^3)$ size circuit, see details in \cite{bartschi2020grover}.

The decision vector $y$ is encoded by an $(\!N\!-\!1\!)$-qubit quantum register.
As $y$ is unconstrained, the corresponding equal superposition can be prepared using Hadamard gates ($H$), as follows:
\begin{align}
    \vert F_{y} \rangle \coloneqq H^{\otimes(N-1)} \vert 0 \rangle^{\otimes(N-1)} = \frac{1}{\sqrt{2^{(N-1)}}}\sum_{y\in F_y} \vert y \rangle.
\end{align}

Thus, we have the operation $U_S$ for preparing an equal amplitude superposition of all feasible states, 
as follows,
\begin{align}
    U_S \coloneqq U_{x}\otimes H^{\otimes(N-1)}.
\end{align}

\begin{algorithm}[h]
    \caption{Condition encoding}
    \label{alg:algorithm2}
    \KwIn{Encoding $x,y$; Demands $q$; Capacity $Q$}
    \KwOut{Conditions $a$}
    
    $c \gets \mathbf{0}$\;
    $d \gets 0$\;
    
    \For{$t \sim  \left[1,2,\dots,N\right]$}{
        \tcc{Log the satisfied demand}
            \For{$i \sim  \left[1,2,\dots,N\right]$}{
                \If{$x_{ti} = 1$}{
                    $d \gets d + q_i$\;
                }
            }
        \tcc{--}
    
        \If{$t \neq 1$}{
            \tcc{Encode the condition}
                \If{$d > Q$ and $y_t = 0$}{
                    $a_t \gets 1$\;
                }
                \If{$y_t = 1$}{
                    $a_t \gets 1$\;
                }
            \tcc{--}
        }
    
        \If{$t \neq 1$ and $t \neq N$}{
            \tcc{Recover register}
                \For{$i \sim  \left[1,2,\dots,N\right]$}{
                    \If{$a_t = 1$ and $c_i = 1$}{
                        $d \gets d - q_i$\;
                    }
                }
                \If{$a_t = 1$}{
                    $c \gets \mathbf{0}$ (requires another register $r$)\;
                }
            \tcc{--}
        }
    
        \If{$t \neq N$}{
            \tcc{Log the satisfied customer}
                \For{$i \sim  \left[1,2,\dots,N\right]$}{
                    \If{$x_{ti} = 1$}{
                        $c_i \gets 1$\;
                    }
                }
            \tcc{--}
        }
    }
    \Return{$a$}\;
\end{algorithm}

\subsection{Phase Separation Operation}
\label{sec:pso}
Unlike typical processing,
a direct derivation of the cost Hamiltonian from the cost function is unavailable in our proposal.
Instead, we introduce additional ancilla bits, denoted as $a$,
to signify the conditions at each time step $t\!\in\!\{2,3,\dots,N\}$, as follows,
\begin{equation}
    a_{t} =
    \begin{cases}
        1 & \text{if $y_t=1$, or capacity insufficiency}, \\
        0 & \text{otherwise}.
    \end{cases}
    \label{eqn:at}
\end{equation}
It enables the reformulation of the cost function to be:
\begin{subequations}
    \begin{align}
        \mathcal{C} (x,a) & \coloneqq \sum_{i=1}^{N}w_{0i}x_{1,i} +  \sum_{i=1}^{N}w_{i0}x_{N\!-\!1,i}\label{eqn:cx1}                                                \\
                          & + \sum_{t=2}^{N}[(1-a_t)\sum_{i,j=1;i\neq j}^{N}w_{ij}x_{t\!-\!1,i}x_{t,j} \label{eqn:cx2} \\
                          & \qquad + a_t\sum_{i,j=1;i\neq j}^{N}(w_{i0}x_{t\!-\!1,i}+w_{0j}x_{t,j})].  \label{eqn:cx3}   
    \end{align}
    \label{eqn:cx}
\end{subequations}
Specifically, the terms in (\ref{eqn:cx1}) compute the distances of the first and last locations transitioning to or from the depot, respectively.
The terms in (\ref{eqn:cx2}) and (\ref{eqn:cx3}), respectively, represent the distances of two possible routes that the vehicle travels intermediate pair of locations under two conditions, 
according to Equation (\ref{eqn:at}).

We replace all binary variables belonging to $x$ and $a$ by $(I\!-\!Z)/2$ to obtain the cost Hamiltonian $H_C$ that satisfies 
$\langle x,a \vert H_{\mathcal{C}} \vert x,a \rangle = \mathcal{C} (x,a)$. 
The ancilla qubits in register $a$ are entangled with the decision qubits in register $x,y$ through the unitary operation $U_E$. 
As we further discussed in Section \ref{sec:ceo}, operation $U_E$ serves to store depot visit conditions for each time step into register $a$, 
aligning it with the criteria set by Equation (\ref{eqn:at}).
The phase separation operation is formulated as follows,
\begin{align}
    U_P(\gamma) \coloneqq U_{E}^{\dagger} e^{-i\gamma H_{\mathcal{C}}} U_{E},
\end{align}
where $e^{-i\gamma H_{\mathcal{C}}}$ can be decomposed into $\mathcal{O}(N^2)$ number of CNOT and $z$-axis rotation gates \cite{seeley2012bravyi}.

\subsection{Condition Encoding Operation}
\label{sec:ceo}
To encode conditions into register $a$,
we introduce another three ancilla registers: $d$, $c$, and $r$.
The register $d$ and $c$, respectively containing $\left\lceil \log_{2}(Q\!+\!\max(q)\!+\!1) \right\rceil $ and $N$ qubits,
serve to track the satisfied demands and customers in the sub-route.
The register $r$ is used to recover register $c$.

As the illustrative circuit shown in Figure~\ref{fig:condition_encoding}, 
the condition encoding operation is constructed under the process outlined in Algorithm~\ref{alg:algorithm2}, 
employing 4 procedures as discussed in detail in the followings:

\begin{figure*}[h]
    \includegraphics[width=\textwidth]{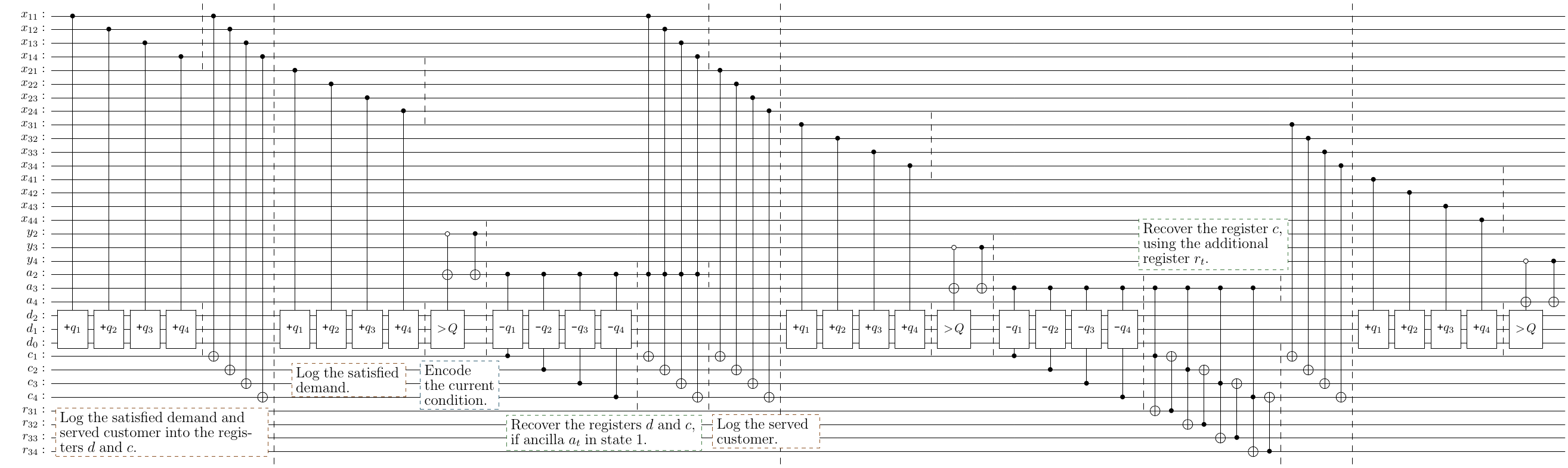}
    \caption{The circuit of condition encoding operation $U_E$. 
    The depot visit condition for each time step is stored into the ancilla register $a$, 
    segmented by full-row barriers. 
    Each step consists of four typical procedures separated by partial-row barriers. 
    First, the satisfied demand of the customer is logged into register $d$. 
    Then, register $d$ and decision qubit $y_t$ control the flip of $a_t$, 
    where $a_t\!=\!1$ suggests a depot visit. 
    After that, if such a depot visit occurs, the register $d$ and $c$ are recovered. 
    Finally, the visited customer number is recorded into register $c$.}\label{fig:condition_encoding}
\end{figure*}

\subsubsection{Log the satisfied demand}
At each time step $t$,
we initially augment register $d$ with the demand of the customer currently being visited.
This operation is carried out by employing $N$ number of $x_{t,i}$-controlled $\left[+q_i\right]$ gates,
for all $i\!\in\!\{1,2,\dots,N\}$.
As $x_t$ preserves hamming weight of 1,
only one $\left[+q_i\right]$ gate can be activated to add the satisfied demand to $d$.
The $\left[+q_i\right]$ gate works as follows,
\begin{align}
    \left[+q_i\right]_{n} \vert d \rangle \coloneqq \vert (d+q_i)\ \mathrm{mod}\ 2^n\rangle,
\end{align}
where $n$ is the number of qubits acting on and $d$ encodes the satisfied demand in binary.

Suppose $\tilde{q}_{i}\!=\!\tilde{q}_{i,K-1}\dots\tilde{q}_{i,0}$ is a binary representation of $q_i$,
we can decompose the $\left[+q_i\right]_{n}$ gate as,
\begin{gather}
    \begin{split}
        \left[+q_i\right]_{n} &= \prod_{k=0,\tilde{q}_{i,k}=1}^{K-1}\left[+2^{k}\right]_{n}\\
        &=\prod_{k=0,\tilde{q}_{i,k}=1}^{K-1}\begin{bmatrix}
            0           & I_{2^k} \\
            I_{2^n-2^k} & 0
        \end{bmatrix}\\
        & =\prod_{k=0,\tilde{q}_{i,k}=1}^{K-1}\left[+1\right]_{n-k} \otimes I_{2^k},
    \end{split}
\end{gather}
where $I_{2^k}$ denotes a Identity matrix of size $2^k$.
\cite{Gidney2015} exhibits a construction that turns $\left[+1\right]_n$ into $\mathcal{O}(n)$ Toffoli, CNOT, and Pauli-$X$ gates.
The number of $\left[+1\right]$ operations required to generate $\left[+q_i\right]$ is tied to the count of `1' present in $\tilde{q}_{i}$.
As $q_i\leqslant Q$, it is restricted to $\mathcal{O}(\log_2Q)$.
Hence, this logging operation requires $\mathcal{O}(N(\log_2Q)^2)$ gates.

\subsubsection{Encode the condition}
The encoding of the condition at time step $t$ is implemented following the logging of satisfied demand.
As depicted in Equation~(\ref{eqn:at}), $a_t$ turns to $1$ when $y_t = 1$ or in the case of capacity insufficiency.
Hence, our first operation involves the deployment of a CNOT gate,
designed to flip the state of $a_t$ when $y_t$ is in state $\vert 1 \rangle$.

The register $d$ encodes the sum of the current demand and demands previously satisfied,
implying a depot visit when the encoded number exceeds $Q$.
To account for all distinct circumstances where $d$ is larger than $Q$,
we employ a series of multi-controlled Pauli-$X$ gates to flip $a_t$,
denoted as,
\begin{align}
    \begin{split}
        &\left[>\!\displaystyle Q \right] \coloneqq \prod_{i=0,\tilde{Q}_i=0}^{K-1} MCX_{\tilde{Q}_{K\!-\!1}\dots\tilde{Q}_{i+1}1}^{d_{K\!-\!1}d_{K\!-\!2}\dots d_{i}}, \\
        & ( K = \left\lceil \log_{2}(Q\!+\!\max(q)\!+\!1) \right\rceil) ,
    \end{split}
    \label{eqn:dq}
\end{align}
where $\tilde{Q}$ is a binary representation of $Q$. 
$MCX_{cs}^{cq}$ denotes the multi-controlled Pauli-$X$ gate, with $cq$ as the control qubits and $cs$ as the control states of $cq$.
A demonstrative example of $\left[>\!\displaystyle Q \right]$ is provided in Figure~\ref{fig:delta_q}.
$\left[>\!\displaystyle Q \right]$ is controlled by $y_t$, and can only be activated when $y_t$ is in state $\vert 0 \rangle$.
It prevents a double flip of $a_t$ when capacity insufficiency occurs and $y_t$ is in state $\vert 1 \rangle$.

An $n$-controlled Pauli-$X$ gate can be realized via $\mathcal{O}(n)$ Toffoli, CNOT, and single-qubit gates.
Thus, we can implement $\left[>\!\displaystyle Q\right] $ within $\mathcal{O}((\log_2Q)^2)$ size circuit.

\begin{figure}[t]
    \centering
    \includegraphics[width=60mm]{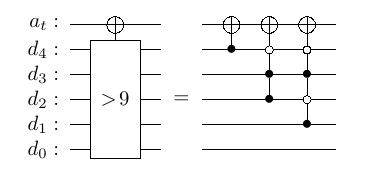}
    \caption{The decomposition of $\left[>\!\displaystyle Q \right]$, where $Q\!=\!9$,
        can be presented in a binary format as `01001'.
        According to Equation~(\ref{eqn:dq}),
        the $\left[>\!\displaystyle 9 \right]$ can be decomposed into a sequence of three multi-controlled Pauli-$X$ gates.
        Specifically, the control states for each gate are identified as `1', `011', and `0101',
        each, respectively, corresponding to control qubits $d_4$, $d_4d_3d_2$, and $d_4d_3d_2d_1$.}\label{fig:delta_q}
\end{figure}

\begin{figure*}[tb]
    \centering
    \includegraphics[width=10cm]{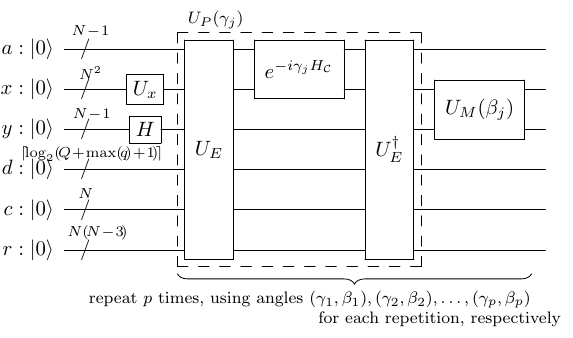}
    \caption{The overall AOA circuit of the CVRP for preparing the state $\vert \psi_p(\boldsymbol{\gamma},\boldsymbol{\beta}) \rangle$.
    $x$ and $y$ are the decision registers, 
    supported by the ancilla registers $a$, $d$, $c$, and $r$. 
    $U_x$ and $H$ initialize the equal amplitude superposition of all feasible states. 
    Then, the operations $U_P(\gamma_j)$ and $U_M(\beta_j)$ are repeated $p$ times, 
    where the mixing operation $U_M(\beta_j)$ is optional. 
    It can be selected from alternatives such as Partialized Mixer $U^{(PM)}_M(\beta_j)$, 
    Grover Mixer $U^{(GM)}_M(\beta_j)$, or other mixers that can preserve the constraints of register $x$.
    Parameters $\boldsymbol{\gamma}$ and $\boldsymbol{\beta}$ are tuned by minimizing the expected value 
    $E_p(\boldsymbol{\gamma},\boldsymbol{\beta}) \coloneqq \langle\psi_p(\boldsymbol{\gamma},\boldsymbol{\beta}) \vert U_E^{\dagger} H_{\mathcal{C}}U_E \vert \psi_p(\boldsymbol{\gamma},\boldsymbol{\beta}) \rangle$.
    }\label{fig:circuit}
\end{figure*}

\subsubsection{Recover the registers and log the satisfied customer number}
Here, we discuss the last 2 procedures. Qubit $a_t$, in $\vert 1 \rangle$ state,
signals a depot visit,
signifying the starting of a new sub-route at the time step $t$.
In this case, the satisfied demand should only encompass the last recording.
As the customers who have been served are monitored using register $c$,
that $c_i$ being in state $\vert 1 \rangle$ indicates the customer $i$ has been visited,
we can employ $c_i$- and $a_t$-controlled operations $\left[-q_i\right]$ to isolate the current demand. 
$\left[-q_i\right]$ is the inverse operation of $\left[+q_i\right]$:
\begin{align}
    \left[-q_i\right] = \left[+q_i\right]^\dagger,
\end{align}
which allows us to minus the demands from the previous sub-route encoded in register $d$.
Simultaneously, this operation ensures that the initial state of $d$ in the subsequent time step encodes a number that does not surpass $Q$.
Hence, $d+q_i$ is confined to the range $[\min(q), Q+\max(q)]$,
and $d$ requires $\left\lceil \log_{2}(Q\!+\!\max(q)\!+\!1) \right\rceil$ qubits.

Register $c$ is recovered back to $\vert 0 \rangle$ states,
at the beginning of a new sub-route.
Two approaches for executing this recovery are presented in Figure~\ref{fig:condition_encoding}.
At time step $2$, $c$ only logs the customer number pertaining to the visit at time step $1$.
Therefore, $x_{1,i}$ share the same state as $c_i$.
We can recover $c$ through the application of Toffoli gates,
utilizing $x_1$ and $a_2$ as controllers.
For time step $3$ to $N\!-\!1$,
we introduce another register $r_t$ for recovery.
A Toffoli gate is applied on $c_i$, $a_t$, and $r_{t,i}$,
followed by a CNOT gate on $c_i$ and $r_{t,i}$,
where $r_{t,i}$ serves as the target and control qubit in two operations, respectively.
Following these operations, $c_i$ is in state $\vert 0 \rangle$, regardless of its previous state, when $a_t$ is $\vert 1 \rangle$.

After completing the recovery,
the current satisfied customer number is logged into register $c$ using CNOT gates.
This guarantees the conservation of the current satisfied demand in register $d$.

It should be noted that during the last time step,
these 2 procedures are unnecessary as the encoding process is finished.
Consequently, $r$ comprises $N\!-\!3$ number of $N$-qubit register.
Recovering register $d$ requires $\mathcal{O}(N (\log_2Q)^2)$ gates, 
which is the same as when logging to $d$, while register $c$ requires $\mathcal{O}(N)$ gates.

\begin{figure*}[tb]
    \centering     
    \subfigure[$P1$: $Q\!=\!3$, $V\!=\!\{(0.66,0.41),\{(0.67,0.23),1\},\{(0.81,0.64),2\},\{(0.17,0.26),2\},\{(0.92,0.46),1\}\}$.]{\label{fig:fig5a}\includegraphics[width=\textwidth]{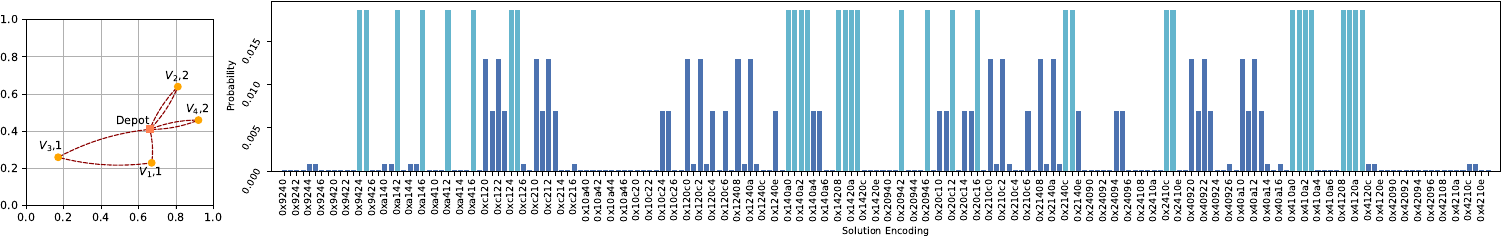}}
    \subfigure[$P2$: $Q\!=\!4$, $V\!=\!\{(0.05,0.68),\{(0.80,0.80),1\},\{(0.97,0.44),3\},\{(0.83,0.25),1\},\{(0.05,0.49),2\}\}$.]{\label{fig:fig5b}\includegraphics[width=\textwidth]{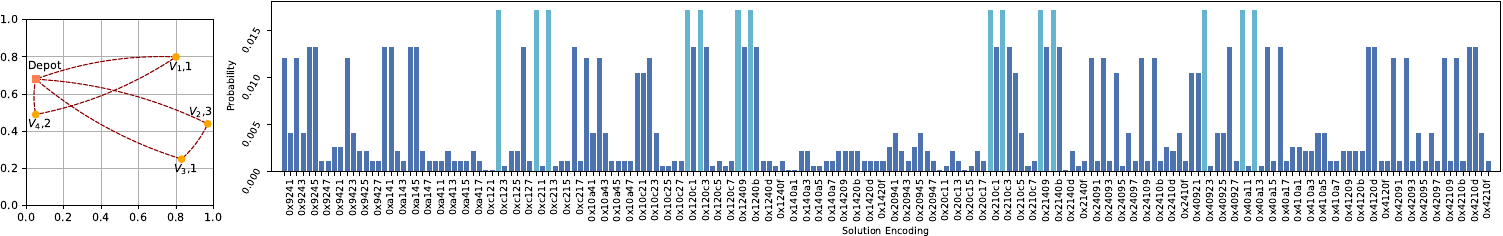}}
    \caption{Two different problems \subref{fig:fig5a} and \subref{fig:fig5b}, with the optimal solutions (left),
        and the measured probability distributions of prepared states using depth-$1$ circuits (right).
        The bars in cyan denote the probabilities of measured optimal encodings.
        Due to the property of the Grover Mixer, the encodings corresponding to the same objective value share the same measured probability.}\label{fig:p1p2}
\end{figure*}

\subsection{Mixing Operation}
\label{sec:mixer}
The mixing operation acts on the register $x$ and $y$. 
It is required to preserve the state of register $x$ as a superposition composed of permutation matrix encodings, 
with the encoding of every single state belonging to the set $F_x$. 
We discuss two possible applicable mixers under this requirement as following.
\subsubsection{Partialized Mixer}
\cite{hadfield2019quantum} introduces a mixer specifically designed for permutation matrix encoding. 
The Hamiltonian of this mixer is the aggregate of the same constructed partial Hamiltonian. 
The partial Hamiltonian is designed to satisfy the swapping between two rows of the permutation matrix that 
\begin{equation}
    H_{PS,\{i,j\}}\vert x_i, x_j \rangle = \vert x_i, x_j \rangle, \; \forall i,j\in\{1,2,\dots,N\},i\neq j.
\end{equation}
It is defined as,
\begin{equation}
    \begin{aligned}
        H_{PS,\{i,j\}} & \coloneqq \sum_{u=1}^{N-1}\sum_{v=u+1}^{N} H_{S,\{i,j\},\{u,v\}},\\
        H_{S,\{i,j\},\{u,v\}} & \coloneqq S^{+}_{x_{iu}}S^{+}_{x_{jv}}S^{-}_{x_{iv}}S^{-}_{x_{ju}}+S^{-}_{x_{iu}}S^{-}_{x_{jv}}S^{+}_{x_{iv}}S^{+}_{x_{ju}},\\
        S^{+} & \coloneqq X+iY, \quad S^{-} \coloneqq X-iY,
    \end{aligned}
\end{equation}
where $X$ and $Y$ denote the Pauli-$X$ and Pauli-$Y$, respectively.
Similar to the XY-Mixer used for one-hot encoding as discussed in \cite{wang2020x}, 
the partial Hamiltonian can also construct complete-graph and ring mixers as follows, 
\begin{equation}
    \begin{aligned}
        H_{complete} & \coloneqq \sum_{i=1}^{N-1}\sum_{j=i+1}^{N} H_{PS,\{i,j\}},\\
        H_{ring} & \coloneqq \sum_{i=1}^{N} H_{PS,\{i,i+1\}}.
    \end{aligned}
\end{equation}
Here, we opt for the ring mixer for the register $x$. 
As the register $y$ remains unconstrained, 
we straightforwardly adopt the X-mixer. 
Finally, the mixer is defined,
\begin{equation}
    \begin{aligned}
        & U_M^{(PM)}(\beta) \coloneqq e^{-i\beta H_{ring}}\otimes e^{-i \beta \sum_{t=2}^{N}X_{y_t}}\\
        & \thickapprox \left(\prod_{i=1}^{N}\prod_{u=1}^{N-1}\prod_{v=u+1}^{N}e^{-i\beta H_{S,\{i,i\!+\!1\},\{u,v\}}}\right)\left(\prod_{t=2}^{N}e^{-i\beta X_{y_t}}\right),
    \end{aligned}
\end{equation}
requiring $\mathcal{O}(N^3)$ basic gates for implementation

\subsubsection{Grover Mixer}
\cite{bartschi2020grover} provides a general form of mixer that preserves the constraints but requires a preparation unitary for the equal superposition of all feasible states, 
known as the Grover Mixer. 
The Hamiltonian of the Grover Mixer is defined as $\vert F \rangle\langle F \vert$. 
Thus, the mixing operation has a Grover-like form~\cite{grover1996fast}, 
\begin{align}
    \begin{split}
        U_{M}^{(GM)}(\beta) & \coloneqq e^{-i\beta \vert F \rangle\langle F \vert} \\
        & = U_S (I-(1-e^{-i\beta})\vert 0 \rangle\langle 0 \vert ^{\otimes n})U_S^{\dagger},
    \end{split}
\end{align}
which is composed of $\mathcal{O}(U_S + n)$ gates, where $n$ is the qubits number. 
For solving the CVRP, the mixer acts on the register $x$ and $y$, requiring $\mathcal{O}(N^3)$ basic gates.

As proven in \cite{bartschi2020grover}, the Grover Mixer restricts the search space to $F$, 
and the states that correspond to an identical objective value have the same amplitude.

\subsection{Resource Evaluation}
Figure~\ref{fig:circuit} shows the end-product circuit that is composed of the operations discussed in this entire section. 
The circuit require 6 quantum register sets as $x$, $y$, $a$, $d$, $c$, and $r$,
involving
\begin{align}
    2N^2-\left\lceil \log_{2}(Q\!+\!\max(q)\!+\!1) \right\rceil-2
\end{align}
qubits in total.
The quantum state evolves via $p$ pairs of phase separation and mixing operation,
whose costs are, respectively, 
$\mathcal{O}(N^2(\log_2Q)^2)$ and $\mathcal{O}(N^3)$, after decomposing into Toffoli, CNOT, and single-qubit gates.

\section{Experiments}
\label{sec:experiments}
\subsection{Experiment Environment and Evaluation Metrics}

We choose the Grover Mixer for the mixing operation and measure the register $x$ and $y$ directly.
We employ Qiskit's statevector simulator \cite{Qiskit} to perform the quantum circuit simulation,
which provides a noise-free probability distribution of measured states when calculating the expected value.
Concurrently, the Constrained Optimization BY Linear Approximation (COBYLA) optimizer \cite{powell1994direct} is adopted to find appropriate parameters for circuits.

Due to the limitation of the simulation environment, we can only construct simple problems.
Here, we mainly focus on the convergence of the prepared state,
ideally reaching the ground state of the cost Hamiltonian ($H_C$).
We propose to evaluate the convergence using the \emph{optimality gap}:
computed as,
\begin{align}
    \label{eqn:opt_gap}
    \alpha \coloneqq \frac{E_p(\boldsymbol{\gamma},\boldsymbol{\beta})}{C(f^{\ast})}-1,
\end{align}
where $f^{\ast}$ encodes an optimal solution that corresponds to the minimal cost value.
It signifies the energy difference between the prepared state and the ground state,
a lower value of optimality gap corresponds to a more desirable outcome.
Furthermore, another metric \emph{optimality ratio}, 
\begin{align}
    r_{opt} \coloneqq \frac{\text{\# optimal shots}}{\text{\# shots}},
\end{align}
is introduced.
The statevector simulator allows the direct computation of $r_{opt}$ as,
$\sum_{f^{\ast}} \left\lvert \langle f^{\ast} \vert  \psi_p(\boldsymbol{\gamma},\boldsymbol{\beta})  \rangle\right\rvert ^2 $.
This metric indicates the probability of measuring the optimal solution,
which directly affects the performance of the solver in real practice.
For the results from QAOA, we extend our evaluation to include the \emph{feasibility ratio}:
\begin{align}
    r_{feas} \coloneqq \frac{\text{\# feasible shots}}{\text{\# shots}}.
\end{align}
Here, it is calculated as 
$\sum_{f \in F} \left\lvert \langle f \vert  \psi_p(\boldsymbol{\gamma},\boldsymbol{\beta})  \rangle\right\rvert ^2 $.

\begin{figure}[t]
    \centering
    \includegraphics[width=80mm]{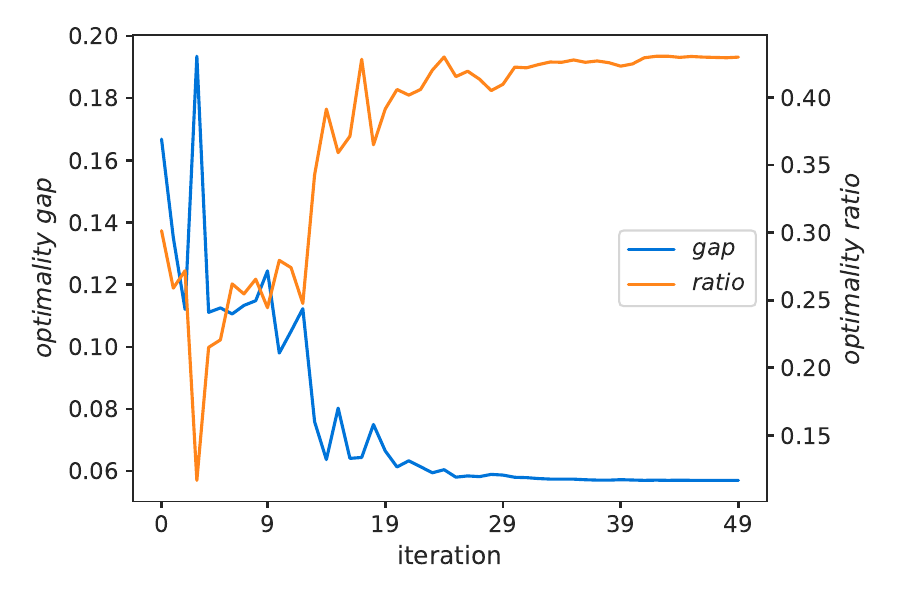}
    \caption{The optimality gap and the optimality ratio of depth-$2$ circuit for $P2$ during the optimization.}\label{fig:p2depth2}
\end{figure}

\begin{figure}[t]
    \centering
    \includegraphics[width=80mm]{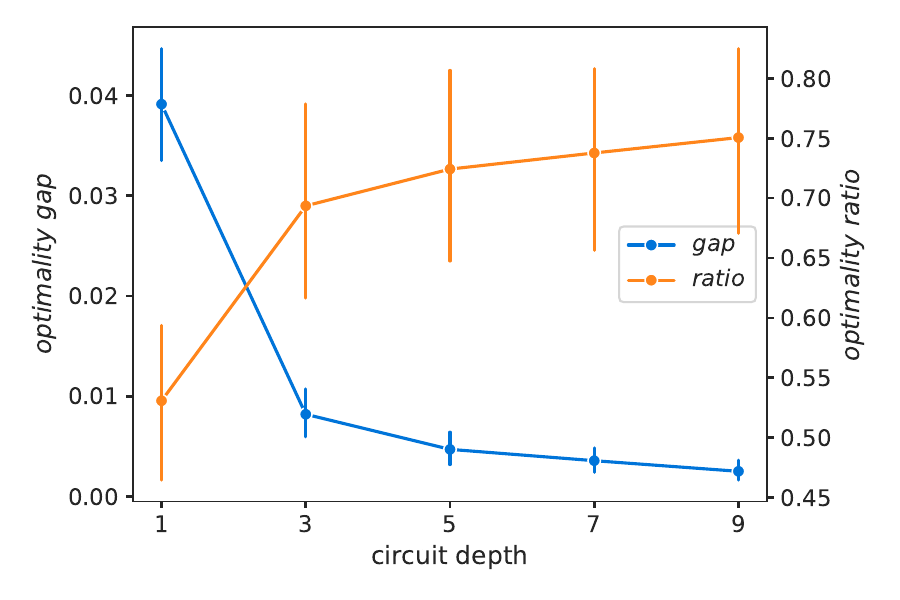}
    \caption{The optimality gap and the optimality ratio of depth-$1$ to depth-$9$ circuits for $P3s$.}\label{fig:p3sresults}
\end{figure}

\subsection{Results and Discussion}
We begin by building two distinct instances of a $4$-customer problem,
denoted as $P1$ and $P2$.
We simply assume that the vehicle is permitted to travel between any pair of locations with a bidirectional, consistent distance.
Notably, the optimal solutions for $P1$ and $P2$, respectively, consist of $3$ and $2$ routes.

We initially solve the problems with the depth-$1$ circuits.
Figure~\ref{fig:p1p2} shows the measured probability distribution of the optimized circuits for both problem instances.
In the case of $P1$, the optimal encodings dominate the distribution, resulting in a significant optimality ratio of $0.597$.
Meanwhile, the optimality gap being $0.013$ also indicates that the circuit of $P1$ converges well.
However, in $P2$, although the probability of optimal encodings is the highest,
the optimality ratio is only $0.241$. This falls below the sum of the secondary peak probabilities.
The reason for this due to the discrepancy in the number of encodings associated with the shortest total path in the feasible set.
While the number of optimal encodings in $P2$ is $14$, the number of encodings corresponding to the subsequent highest probability is $23$.
To improve the results,
we apply a depth-$2$ circuit, observing enhanced performance even in the early optimization stage.
As illustrated in Figure~\ref{fig:p2depth2}, the optimality surpasses $0.35$ in the middle, ultimately reaching $0.43$.

We expand the investigation to include deeper circuits on $48$ randomly generated $3$-customer problem instances, $P3s$. 
As circuit depth increases, the optimality gap approaches zero while the optimality ratio increases, as shown in Figure~\ref{fig:p3sresults}.

Compared to the QAOA solver with QUBO encoding, our proposed method demonstrates an entirely different order of magnitude enhancement across all evaluative metrics,
as shown in Figure~\ref{fig:comparison_results} and Table \ref{table1}. 
Attributed to our encoding, the search space can be narrowed to the feasible set, 
thereby vastly reducing the universe of potential solutions when compared to penalty-inclusive QUBO encoding with the conventional QAOA approach. 
Consistent with the conclusions drawn in \cite{azad2022solving,palackal2023quantum},
it is less reliable to use QAOA for solving constrained combinatorial problems.
Furthermore, as the route number is predefined as $\lceil \sum_{i=1}^{N\!-\!1}q_i /Q\rceil $, 
the search space of QUBO for $P1$ fails to cover the optimal solution.

\begin{figure*}[t]
    \centering
    \includegraphics[width=0.8\textwidth]{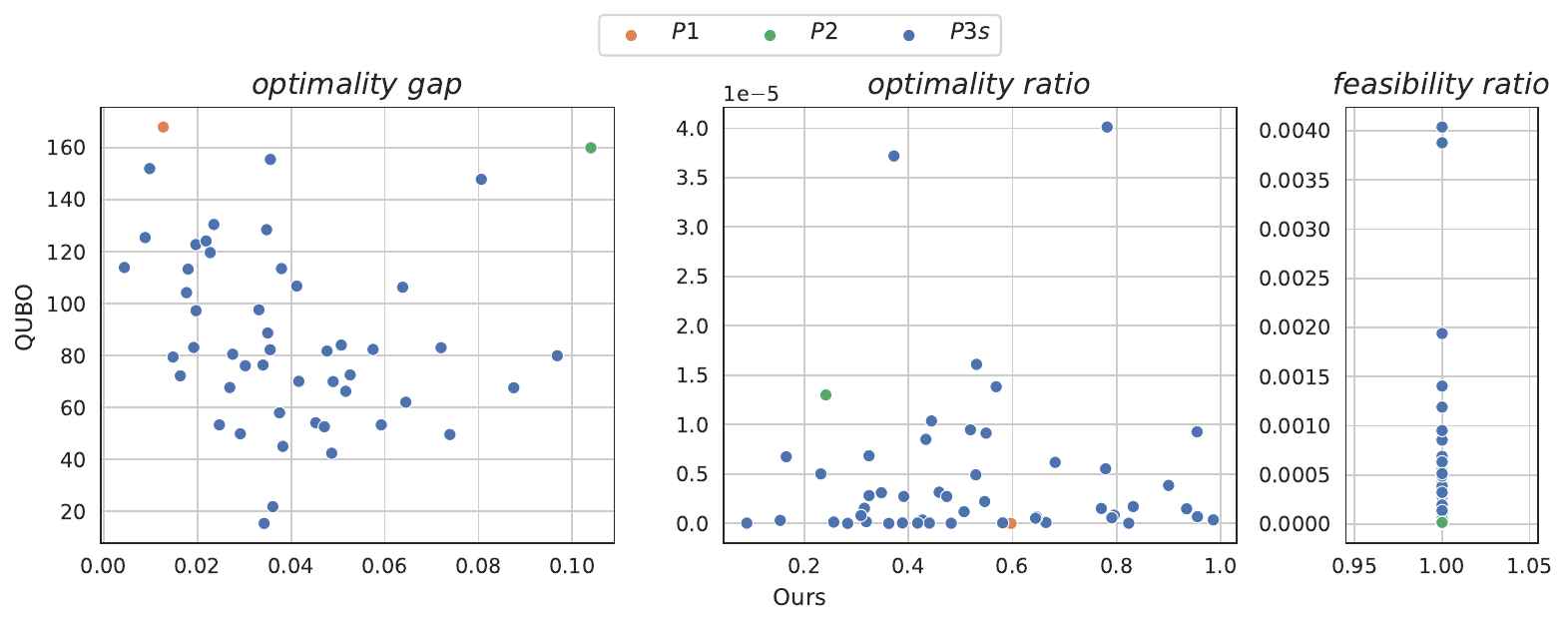}
    \caption{The evaluation results for the optimized depth-$1$ circuits with our and QUBO encoding.}\label{fig:comparison_results}
\end{figure*}

\begin{table}[t]
    \centering
    \caption{The evaluation results (avg.) for the optimized depth-$1$ circuits with different problem encoding.}\label{table1}
    \setlength{\tabcolsep}{2pt}
    \renewcommand{\arraystretch}{1.5}
    \resizebox{\columnwidth}{!}{
    \begin{tabular}{ccccccc}
    \toprule
                       & \multicolumn{2}{c}{$P1$}      & \multicolumn{2}{c}{$P2$}      & \multicolumn{2}{c}{$P3s$}      \\ \cline{2-7} 
      Encoding  & Ours      & QUBO          & Ours       & QUBO          & Ours       & QUBO          \\ \midrule
    \emph{optimality gap}    & $1.27e\!-\!2$ & $1.68e\!+\!2$ & $1.04e\!-\!1$ & $1.60e\!+\!2$ & $3.91e\!-\!2$ & $8.50e\!+\!2$ \\
    \emph{feasibility ratio} & $1$           & $1.77e\!-\!5$ & $1$           & $1.78e\!-\!5$ & $1$           & $5.76e\!-\!4$ \\
    \emph{optimality ratio}  & $5.97e\!-\!1$ & $0$           & $2.41e\!-\!1$ & $1.30e\!-\!5$ & $5.31e\!-\!1$ & $4.65e\!-\!6$ \\ \bottomrule
    \end{tabular}}
    \renewcommand{\arraystretch}{1}
\end{table}

\section{Conclusion}
\label{sec:conclusion}
Limiting the search space in the feasible subspace and initializing from the equal amplitude superposition of all feasible states have been numerically proven to enhance the solution quality for QAOAs when solving constrained combinatorial optimization problems \cite{wang2020x}. 
However, the applicable range of problems is limited, 
as the current families of mixing and preparation operations are suitable for fewer types of feasible sets. 
Via reformulation of the problem encoding, this challenge might be mitigated.

In this work, we re-encode the CVRP, allowing quantum states to initialize from an equal amplitude superposition of all feasible states, 
and enabling the solver to preserve the customer visit constraint via mixers.
Meanwhile, the vehicle capacity constraint is ensured through conditional decoding. 
Following this, we introduce ancilla conditional indicator bits to derive the cost Hamiltonian.
We further demonstrate a method to encode conditions into the ancilla qubits using a polynomial number of basic gates for the phase separation operation. 
These approaches also carry the potential for extensions to other constrained combinatorial optimization problems, for example, the Knapsack Problems, 
as they have similar problem structures.

Using statevector simulation, we validate that the developed AOA solver indeed preserves feasibility by several illustrative problems. 
Furthermore, we show that the solver outperforms the conventional QAOA approach, 
exhibiting not only higher feasibility ratios, but also higher optimality ratios, and reduced optimality gaps. 
Notably, these metrics show improvements in the order of magnitude, 
emphasizing the effectiveness and potential of our proposal.

Despite the advantages of the AOA framework, 
our proposed method has several drawbacks. 
The designed circuit, although with a polynomial number of gates, 
requires many multi-controlled Toffoli (MCT) gates. 
The MCT gate is not friendly to topological quantum devices, 
as it might need additional SWAP gates to adapt to the topology. 
Also, the bit-flip noises in real quantum devices can easily destroy the feasible space preserved by the mixers. 
Thus, it is questionable whether our proposed method is suitable to run on near-term Noisy Intermediate-Scale Quantum (NISQ) \cite{preskill2018quantum} devices. 
Future work must focus on further optimizations and robustness against quantum noise for practical deployment.

\bibliographystyle{IEEEtran}
\bibliography{gmcvrp}

\end{document}